\documentclass[prl,twocolumn,showpacs,nobalancelastpage,hyperref]{revtex4}%
\usepackage{epsfig}
\usepackage{color}
\usepackage{ulem}
\usepackage{amsmath}
\usepackage{amsfonts}
\usepackage{amssymb}
\usepackage{graphicx}%
\providecommand{\U}[1]{\protect\rule{.1in}{.1in}}
\newcommand{\be}{\begin{equation}}
\newcommand{\ee}{\end{equation}}
\newcommand{\bea}{\begin{eqnarray}}
\newcommand{\eea}{\end{eqnarray}}

\begin{document}
\title{ Spontaneous Emission from a Fractal Vacuum}
\author{Eric Akkermans and Evgeni Gurevich}
\affiliation{Department of Physics, Technion Israel Institute of Technology, Haifa 32000, Israel}

\begin{abstract}
Spontaneous emission of a quantum emitter coupled to a QED vacuum with a
deterministic fractal structure of its spectrum is considered. We show that
the decay probability does not follow a Wigner-Weisskopf exponential decrease
but rather an overall power law behavior with a rich oscillatory structure,
both depending on the local fractal properties of the vacuum spectrum. These
results are obtained by giving first a general perturbative derivation for
short times. Then we propose a simplified model which retains the main
features of a fractal spectrum to establish analytic expressions valid for all
time scales. Finally, we discuss the case of a Fibonacci cavity and its
experimental relevance to observe these results.

\end{abstract}

\date{\today}
\maketitle

%----------------------------------------------------------------------------

%\maketitle
%--------------------------------------------------------------------------

%-----------------------------------------------------------------------

%INTRODUCTION

Spontaneous emission results from the coupling of a quantum system with a
discrete energy spectrum (an "atom") to a quantum vacuum. This is an important
and widely studied phenomenon both from fundamental and applied points of view
\cite{reviews}. Spontaneous emission allows to probe properties of the quantum
vacuum, its dynamics and correlations. The wide zoology of behaviors of
spontaneous emission depends on spectral properties of the vacuum and on its
coupling to the atom \cite{QO-Struct-Vacuum-00}. A standard textbook
description \cite{reviews,CCTbook} considers the coupling to a vacuum having a
smooth and non-singular density of modes of photons, in which case, the
probability for spontaneous emission follows the well-known Wigner-Weisskopf
decay law,%
\begin{equation}
|U_{e}(t)|^{2}=e^{-\Gamma_{e}(\omega_{e})t}\,. \label{WW}%
\end{equation}
Relevant definitions of the quantum amplitude $U_{e}(t)$ and of the inverse
lifetime $\Gamma_{e}(\omega_{e})$ will be given below. This description has
been further developed towards quantum emitters coupled to more complicated
environments such as semiconductors, QED-cavities, photonic crystals and
micro-cavities \cite{QO-Struct-Vacuum-00}. The existence of singularities in
the spectrum of the vacuum leads to a qualitatively different behavior which
has been studied in various cases \cite{kurizki}.

In this letter, we address the problem of spontaneous emission from an atom
coupled to a vacuum whose spectrum is characterized by a \textit{discrete
scaling symmetry} expressed by the property%
\begin{equation}
\mu(\omega+\Delta\omega)-\mu(\omega)={\frac{{\mu}\left(  {T(\omega
+\Delta\omega)}\right)  {-\mu\left(  T\left(  \omega\right)  \right)  }}{a}}
\label{scaling}%
\end{equation}
where $\mu\left(  \omega\right)  $ is the integrated density of modes, or
spectral measure, and the dimensionless scaling parameter $a$ and the map
$T(\omega)$ provide a full characterization of the specific discrete scaling
symmetry. Introducing $\mathcal{N}_{\omega_{u}}\left(  \omega\right)  \equiv$
$\mu(\omega)-\mu\left(  \omega_{u}\right)  $, the scaling relation
(\ref{scaling}) can be written more concisely as $\mathcal{N}_{\omega_{u}%
}\left(  \omega\right)  =\frac{1}{a}\mathcal{N}_{T\left(  \omega_{u}\right)
}\left(  T\left(  \omega\right)  \right)  $. A spectrum described by
(\ref{scaling}) is often called fractal (see \cite{reviewfractals} and refs.
therein), a denomination that we shall retain all over but which covers a
broad class of systems that extends beyond the conventional self-similar
character associated to fractals, such as the popular Cantor set. More
involved examples include cavities made out of quasi-periodic, \textit{e.g.}
Fibonacci, sequences of dielectric layers, or cavities generated from
deterministic self-similar fractal structures like a Sierpinski gasket. The
prominent feature of fractal spectra is that they are highly lacunar,
possessing an infinity of gaps appearing at all scales. This is a direct
consequence of their discrete scaling symmetry (\ref{scaling}). It is worth
noticing that for a singular spectrum, the integrated density of modes
$\mu(\omega)$ is usually well defined, while its derivative, the density of
modes $\rho(\omega)=\frac{d\mu(\omega)}{d\omega}$, may not be.

Singularities of the spectrum satisfying (\ref{scaling}) correspond to fixed
points of the map $T\left(  \omega\right)  $. Around a fixed point $\omega
_{u}$ the map can be linearized, $T(\omega)\simeq\omega_{u}+T^{\prime}%
(\omega_{u})(\omega-\omega_{u})$. Thus, the spectral measure obeys the
equation
%TCIMACRO{\QTR{QQQEvgeni}{ }}%
%BeginExpansion
%EndExpansion
$\mathcal{N}_{\omega_{u}}\left(  \omega\right)  =\frac{1}{a}\mathcal{N}%
_{\omega_{u}}\left(  \omega_{u}+T^{\prime}\left(  \omega_{u}\right)  \left(
\omega-\omega_{u}\right)  \right)  $, whose general solution is
\cite{Fract-Coarse-gran}%
\begin{equation}
{\mathcal{N}_{\omega_{u}}\left(  \omega\right)  =|\omega-\omega_{u}|^{\alpha
}\,\mathcal{F}\left(  {\frac{\ln|\omega-\omega_{u}|}{\ln|T^{\prime}(\omega
_{u})|}}\right)  ,} \label{mu}%
\end{equation}
where $\alpha=\frac{\ln a}{{\ln|T^{\prime}(\omega_{u})|}}$ is the local
($\omega_{u}$-dependent) spectral exponent and $\mathcal{F}(x)$ is a periodic
function of period unity \cite{Note-Ds}. Generically, $\alpha$ changes between
zero and unity spanning the range between smooth continuous and point spectrum
\cite{Bellisard-Transport}.

We show below that the coupling of an atom whose resonance frequency is close
to $\omega_{u}$ leads to a similar scaling behavior of the time dependent
decay amplitude $U_{e}(t)$, given in the long time limit by%
\begin{equation}
\left\vert U_{e}(t)\right\vert ^{2}=t^{-2\gamma}\,\mathcal{G}\left(
{\frac{\ln t}{\lambda}}\right)  ,\label{main}%
\end{equation}
where $\mathcal{G}(x+1)=\mathcal{G}(x)$ is another periodic function and
$\gamma$ is a function of the spectral exponent $\alpha$
\cite{Note-IncompDecay}. The exponent $\gamma$, the real parameter
$\lambda\equiv\ln|T^{\prime}(\omega_{u})|$ and the specific form of the
function $\mathcal{G}(x)$ are direct consequences of the specific scaling
relation (\ref{mu}). A similar scaling has been obtained for the
thermodynamics of a fractal blackbody \cite{ADT2}. Thus, for a fractal
spectrum, spontaneous emission does not follow the Wigner-Weisskopf
exponential decay (\ref{WW}), a result which could be partly anticipated from
the singular nature of the spectrum. It is indeed known that the existence of
spectral singularities leads to an algebraic time decrease of the decay
probability qualitatively different from the exponential behavior or its small
time approximation given by the Fermi golden rule \cite{kurizki}. But the
existence of log-periodic fluctuations described by the function
$\mathcal{G}(x)$ is a direct consequence of the discrete scaling symmetry
(\ref{scaling}).

Expression (\ref{main}) constitutes the main result of this letter. To
establish it, we shall first recall some basic definitions and results. Then,
we will give a
%TCIMACRO{\QTR{QQQEric}{general perturbative} }%
%BeginExpansion
general perturbative
%EndExpansion
derivation starting from the Fermi golden rule, essentially limited to small
times. To go beyond this limit, we will consider a model general enough to
include all relevant characteristics of a fractal spectrum as given by
(\ref{scaling}), yet simple enough to allow for a thorough analytical
derivation. Finally, we will discuss the case of the Fibonacci cavity which
provides a relevant example of quantum vacuum with discrete scaling symmetry
which can be realized experimentally.

%DEFINITIONS AND NOTATIONS
A two-level quantum system $(|g\rangle,|e\rangle)$ whose Hamiltonian is
$H_{e}=\hbar\omega_{e}\left\vert e\right\rangle \left\langle e\right\vert $,
with respective energies $0$ and $\hbar\omega_{e}$, is coupled to the EM field
described by $H_{F}=\hbar\sum_{k}\omega_{k}a_{k}^{\dagger}a_{k}$, where $k$
stands for an appropriate set of quantum numbers. The atom-photon interaction
is described within the rotating wave approximation by the Hamiltonian
$H_{int}=\sum_{k}\left(  V_{k}^{\ast}a_{k}^{\dagger}\left\vert g\right\rangle
\left\langle e\right\vert +h.c.\right)  $ where only resonant and
near-resonant transitions are considered. The matrix element $V_{k}$ denotes
the strength of the coupling, in general dependent on the atom's position. In
the initial state $|e,0_{k}\rangle$, the atom is in the excited state and no
photon is present. We aim to evaluate the probability amplitude $U_{e}(t)$ to
find the quantum system in the initial state a time $t$ after it evolves with
the total Hamiltonian $H=H_{e}+H_{F}+H_{int}$. This amplitude
%, also called vacuum persistence
is defined by%
\begin{equation}
U_{e}(t)=\langle e,0_{k}|\hat{U}(t,0)|e,0_{k}\rangle\,\ . \label{persist}%
\end{equation}
The evolution operator $\hat{U}(t,0)$, written
%with obvious notations,
in terms of the resolvent operator $\hat{G}(z)=1/(z-H)$, is \cite{CCTbook}%
\begin{equation}
\hat{U}(t,0)=\frac{1}{2\pi i}\int_{-\infty}^{+\infty}dEe^{-iEt/\hbar}(\hat
{G}_{+}\left(  E\right)  -\hat{G}_{-}\left(  E\right)  ). \label{U_Green}%
\end{equation}
The matrix element $G_{e}(z)\equiv\langle e,0_{k}|\hat{G}(z)|e,0_{k}%
\rangle=1/(z-E_{e}-\Sigma_{e}(z))$ of the resolvent defines the self-energy
$\Sigma_{e}(\omega\pm i0^{+})=\Delta_{e}(\omega)\mp\frac{i\hbar}{2}\Gamma
_{e}(\omega)$ in terms of two spectral functions, $\Delta_{e}(\omega)$ and
$\Gamma_{e}(\omega)$ which respectively account for the shift and the spectral
width of the atomic energy $E_{e}$.

The spectral function $\Gamma_{e}(\omega)$ is related to the vacuum response
function%
\begin{equation}
\Phi_{e}(t)={\frac{1}{2\pi}}\int d\omega\,\Gamma_{e}(\omega)\,e^{-i\omega
t}\,, \label{phigamma}%
\end{equation}
which can also be expressed using the spectral measure $\mu(\omega)$ as%
\begin{equation}
\Phi_{e}(t)={\frac{1}{\hbar^{2}}}\int d\mu(\omega_{k})|V_{k}|^{2}%
e^{-i\omega_{k}t}. \label{phimu}%
\end{equation}
Within the dipole approximation, $V_{k}=\mathcal{E}_{k}\left(  \mathbf{r}%
\right)  \left(  \hat{e}_{z}\cdot\hat{\varepsilon}_{k}\right)  d_{ge}$, the
response $\Phi_{e}(t)$ is the time correlation function,%
\begin{equation}
\Phi_{e}(t)=\hbar^{-2}\left\vert d_{ge}\right\vert ^{2}\langle0_{k}|\hat
{E}_{z}(\mathbf{r},t)\hat{E}_{z}^{\dagger}(\mathbf{r},0)|0_{k}\rangle,
\end{equation}
of the electric field component $\hat{E}_{z}(\mathbf{r},t)$ along the
polarization direction $\hat{z}$ of the atom. Here, $d_{ge}$ and $\mathbf{r}$
are respectively the dipole matrix element and the position of the atom. A
convenient form of the probability amplitude $U_{e}(t)$ is given in terms of
the Laplace transform $\tilde{\Phi}_{e}(s)$ of the correlation function
\cite{CCTbook}:%
\begin{equation}
U_{e}(t)={\frac{1}{2\pi i}}\int_{a-i\infty}^{a+i\infty}ds{\frac{e^{\left(
s-i\omega_{e}\right)  t}}{s+\tilde{\Phi}_{e}(s-i\omega_{e})}.}
\label{persiscorr}%
\end{equation}
%BASIC RESULTS IN THE THEORY OF SPONTANEOUS EMISSION%
Previous definitions and results summarize the state of the art and allow to
identify two relevant energy scales for the problem of spontaneous emission.
One, $\Gamma_{e}(\omega_{e})\simeq|V_{k}|^{2}\rho\left(  \omega_{e}\right)  $,
is the strength of the coupling between the emitter and the vacuum as defined
from the Hamiltonian $H_{int}$. The second energy scale $\Delta$ is given by
the spectral width of $\Gamma_{e}(\omega)$. Relation between the two allows to
identify a dimensionless coupling parameter%
\begin{equation}
g={\frac{\Gamma_{e}(\omega_{e})}{\Delta}.} \label{g}%
\end{equation}
In the weak coupling limit, $g\ll1$, the quantum amplitude $U_{e}(t)$ is
determined by the pole in (\ref{persiscorr}), given by the approximate
solution $s\approx-\tilde{\Phi}_{e}(-i\omega_{e})=-\frac{1}{2}\Gamma
_{e}(\omega_{e})$. This leads straightforwardly to the Wigner-Weisskopf
exponential decay (\ref{WW}) of the probability amplitude.
%TCIMACRO{\TeXButton{red}{\color{red}} }%
%BeginExpansion
\color{red}
%EndExpansion
%TCIMACRO{\TeXButton{black}{\color{black}} }%
%BeginExpansion
\color{black}
%EndExpansion
At extremely long times, $t\gg\Gamma_{e}^{-1}(\omega_{e})$, this pole
approximation is not valid anymore, even in free space, and both the
probability amplitude $U_{e}(t)$ and the correlation function $\Phi_{e}(t)$,
are dominated by the singularity at the edge $\omega=0$ of the vacuum spectrum
\cite{CCTbook}. For an atom coupled to the $d$-dimensional scalar\ QED vacuum,
$\Gamma_{e}(\omega)\simeq\rho(\omega)\left\vert \mathcal{E}\left(
\omega\right)  \right\vert ^{2}$, where the density of states $\rho
(\omega)\simeq\omega^{d-1}$ and the amplitude of the electric field
$\mathcal{E}\left(  \omega\right)  \simeq\sqrt{\omega}$, so that $\Gamma
_{e}\simeq\omega^{d}$ and $\Phi_{e}(t)\simeq1/t^{d+1}$, and according to
(\ref{persiscorr}), $U_{e}(t)\simeq1/t^{d+1}$. Such singularities have also
been studied in the lifetime of electronic states in quantum mesoscopic metals
\cite{eagm}. In more structured vacuum spectra such as photonic crystals, the
spectral function $\Gamma_{e}(\omega)$ exhibits singularities of the type
$\Gamma_{e}(\omega)=C|\omega-\omega_{u}|^{\alpha-1}\theta(\omega-\omega_{u})$
around certain frequencies $\omega_{u}$ \cite{QO-Struct-Vacuum-00}. In that
case the coupling is strong and the pole approximation is not applicable.
Instead, one obtains a \textit{generalized} exponential decay, limited to
small times well accounted for by the Fermi golden rule, $U_{e}(t)-1\simeq
Ct^{2-\alpha}$ (see also discussion below). At large times, it turns into an
algebraic decrease $U_{e}(t)\simeq t^{-\left(  2-\alpha\right)  }$, possibly
coexisting with a non-decaying component \cite{kurizki}.

\begin{figure}[ptb]%
\centering
\includegraphics[
height=2.1677in,
width=2.8908in
]%
{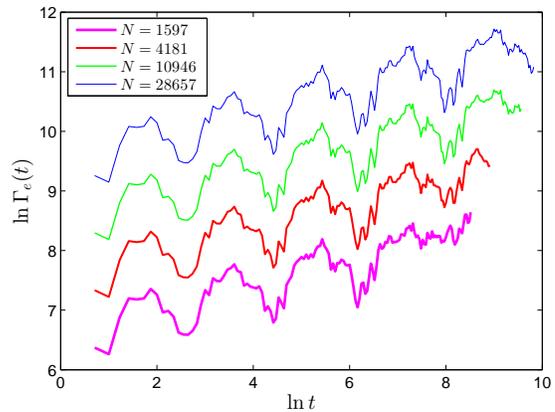}%
\caption{(Color online) Numerical results for $\Gamma_{e}(t)$, given by
(\ref{fgr2}), for Fibonacci dielectric cavities $S_{j}$ [defined in text] of
different lengths $N$. The photonic spectrum has been obtained from a
simplified diagonal tight-binding (TB) description of the Fibonacci potential,
which however exhibits the relevant fractal properties of a dielectric
superlattice. The matrix element $|V_{k}|^{2}$ in (\ref{fgr2}) was taken
constant, which is equivalent to averaging $\Gamma_{e}(t)$ over the atom's
position in the cavity. The dimensionless time $t$ is in units of the TB
hopping constant.}%
\label{fibo}%
\end{figure}

%QUALITATIVE DERIVATION
We now consider the coupling to a fractal vacuum spectrum obeying the scaling
in (\ref{mu}). In this case the weak coupling limit becomes ill-defined, since
the spectral function $\Gamma_{e}(\omega)\simeq\rho\left(  \omega\right)  $ is
a singular function with a vanishing variation scale $\Delta$. Thus, according
to (\ref{g}), we are effectively in a strong coupling regime $g\gg1$, even for
a finite and weak coupling $V_{k}$. On the other hand, the short time
perturbative approach remains applicable. We start with this approximation to
present an intuitively clear explanation of the effect of the vacuum spectrum
fractality on the decay dynamics. At short times one obtains from
(\ref{persiscorr})%
\begin{equation}
\left\vert U_{e}(t)\right\vert ^{2}\simeq1-\int_{0}^{t}dt^{\prime}\,\Gamma
_{e}(t^{\prime}), \label{perturbation}%
\end{equation}
where using (\ref{phigamma}) and the explicit expression of the correlation
function $\Phi_{e}(t)$ we have%
\begin{equation}
\Gamma_{e}(t)={\frac{2t}{\hbar^{2}}}\int d\mu(k)|V_{k}|^{2}\,{\frac
{\sin\left(  \omega_{k}-\omega_{e}\right)  t}{\left(  \omega_{k}-\omega
_{e}\right)  t}.} \label{fgr2}%
\end{equation}
While this expression is textbook materials, we wish to re-examine it in the
context of a fractal spectrum considered here. The $\mbox{sinc}$ function on
the \textit{r.h.s} of (\ref{fgr2}) indicates that $\Gamma_{e}(t)$ is not the
Fourier transform of the spectral function $\Gamma_{e}(\omega)$, but it is
rather a wavelet transform. Generally, the wavelet transform $S_{w}(a,b)$\ of
a function $s(x)$ is defined by \cite{vaienti}%
\begin{equation}
S_{w}(a,b)\equiv{\frac{1}{a}}\int dx\,s(x)\,w\left(  {\frac{x-b}{a}}\right)
\,. \label{wavelet}%
\end{equation}
It can be viewed as a mathematical microscope which probes the function $s(x)$
at a point $b$ with a magnification $1/a$ and an optics specified by the
choice of the specific wavelet $w(x)$. Thus, $\Gamma_{e}(t)$ in (\ref{fgr2})
is the wavelet transform of the spectral measure $\mu(\omega)$ at a frequency
$\omega_{e}$, with a magnification $t$ and $w(x)=\mbox{sinc}(x)$ as a probe.
An important property of the wavelet transform is that it preserves a discrete
scaling symmetry (\ref{scaling}) of the probed function, here the spectral
measure $\mu(\omega)$ weighted by $\left\vert V_{k}\right\vert ^{2}$. For a
smooth and continuous spectrum, the $\operatorname{sinc}$ function probes the
spectrum on energy scales of the order of $t^{-1}$ and in the long time limit
it goes over $\delta(\omega_{k}-\omega_{e})$ so that $\Gamma_{e}(t)$ becomes
$t$-independent as expected. However there is no such well-defined limit for a
fractal spectrum and inserting (\ref{mu}) into (\ref{fgr2}) for $\omega
_{e}=\omega_{u}$, we obtain instead \cite{Note-CouplDOM}%
\begin{equation}
\Gamma_{e}(t)=t^{1-\alpha}\,\tilde{\mathcal{F}_{1}}\left(  {\frac{\ln t}%
{\ln|T^{\prime}(\omega_{u})|}}\right)  \label{gammat}%
\end{equation}
and%
\begin{equation}
\left\vert U_{e}(t)\right\vert ^{2}=1-t^{2-\alpha}\,\tilde{\mathcal{F}_{2}%
}\left(  {\frac{\ln t}{\ln|T^{\prime}(\omega_{u})|}}\right)  ,
\label{Prob_Fermi}%
\end{equation}
where $\tilde{\mathcal{F}}_{1,2}(x)$ are periodic functions of their argument
of period unity. This constitutes a short-time counterpart to the asymptotic
result\ (\ref{main}). The behavior of $\Gamma_{e}(t)$ is illustrated in Fig.
\ref{fibo} for the specific case of a Fibonacci quasi-periodic dielectric
cavity (to be discussed later on), whose spectral measure admits locally a
discrete scaling symmetry (\ref{mu}). We observe, as predicted by
(\ref{gammat}), an overall power law behavior with $\alpha\approx0.8$,
explained by the renormalization group analysis of \cite{Kohmoto-87}, and also
log-periodic oscillations around it, which are the fingerprint of the
underlying fractal structure of the spectrum. Note that the log-periodic
oscillations characteristic of a fractal spectrum are already noticeable for
systems of finite size and remain unchanged as length increases.%
%TCIMACRO{\FRAME{ftbpFU}{2.8908in}{2.1677in}{0pt}{\Qcb{(Color online) Numerical
%results for $\Gamma_{e}(t)$, given by (\ref{fgr2}), for Fibonacci dielectric
%cavities $S_{j}$ [defined in text] of different lengths $N$. The photonic
%spectrum has been obtained from a simplified diagonal tight-binding (TB)
%description of the Fibonacci potential, which however exhibits the relevant
%fractal properties of a dielectric superlattice. The matrix element
%$|V_{k}|^{2}$ in (\ref{fgr2}) was taken constant, which is equivalent to
%averaging $\Gamma_{e}(t)$ over the atom's position in the cavity. The
%dimensionless time $t$ is in units of the TB hopping constant.}}{\Qlb{fibo}%
%}{fibo1.eps}{\special{ language "Scientific Word";  type "GRAPHIC";
%width 2.8908in;  height 2.1677in;  depth 0pt;  original-width 5.5625in;
%original-height 4.1702in;  cropleft "0";  croptop "1";  cropright "1";
%cropbottom "0";  filename 'fibo1.eps';file-properties "XNPEU";}} }%
%BeginExpansion
%EndExpansion

\begin{figure}[ptb]%
\centering
\includegraphics[
height=2.1677in,
width=2.8908in
]%
{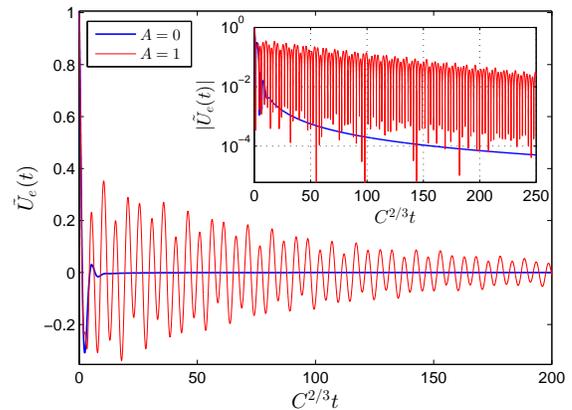}%
\caption{(Color online) Amplitude $\tilde{U}_{e}(t)$ as a function of time for
the model (\ref{toy}) with $\alpha=\frac{1}{2}$, \ $\lambda=1$ and $A=0$ (blue
thick line) or $A=1$ (red thin line). Inset shows $\left\vert \tilde{U}%
_{e}(t)\right\vert $ in a semi-log scale.}%
\label{utoyfig1}%
\end{figure}

%QUANTITATIVE DERIVATION
The previous derivation is based on the Fermi golden rule valid at short
times. In order to go beyond this regime, a more elaborate derivation is
needed. To that purpose, we consider the following model for the spectral
function:%
\begin{equation}
\Gamma_{e}(\omega)=\frac{C}{\pi\left\vert \omega-\omega_{u}\right\vert
^{1-\alpha}}\left[  1+A\cos\left(  \frac{2\pi}{\lambda}\ln\frac{\left\vert
\omega-\omega_{u}\right\vert }{\Omega}\right)  \right]  , \label{toy}%
\end{equation}
where $0<\alpha<1$ is the local spectral exponent introduced in (\ref{mu}),
$C$ is the coupling strength, and\ $0\leq A\leq1$ and $\Omega$ define the
modulation amplitude and phase respectively. For simplicity, we let
$-\infty<\omega<\infty$. This expression exhibits the basic features required
to describe a fractal structure of the vacuum as defined in (\ref{scaling})
and (\ref{mu}) with $\lambda=\ln\left\vert T^{\prime}\left(  \omega
_{u}\right)  \right\vert $. We have approximated the log-periodic function by
its first harmonic, which happens to be a good approximation as shown in
related situations \cite{ADT1, ADT3}. In the absence of log-periodic
modulation, i.e. for $A=0$, we recover the case of a singularity in the
spectrum similar to that already studied in the literature \cite{kurizki}. The
important point here is that the modulation results from the specific scaling
properties of the spectrum defined in (\ref{scaling}) and discussed
subsequently. The model (\ref{toy}) allows to obtain closed analytical
expressions of the quantum amplitude $U_{e}(t)$ and therefore to study its
behavior both in the short time limit (in a way very similar to the discussion
which follows (\ref{perturbation})) and in the long time limit which was not
possible starting from the Fermi golden rule. From (\ref{phigamma}), we obtain%
\begin{equation}
\Phi_{e}(t)=C{\frac{2e^{-i\omega_{u}t}}{\pi t^{\alpha}}}\left[  \Gamma
(\alpha)\cos{\frac{\pi\alpha}{2}}-A\,\mbox{Im}\,F(t)\right]  , \label{phittoy}%
\end{equation}
where $\Gamma(x)$ is the Euler Gamma-function and we have defined
$F(t)\equiv\left(  \Omega t\right)  ^{2i\pi/\lambda}\,\cosh\left(  \frac
{\pi^{2}}{\lambda}+\frac{i\pi\alpha}{2}\right)  \,\Gamma\left(  \alpha
-{\frac{2i\pi}{\lambda}}\right)  $. We thus immediately notice that $\Phi
_{e}(t)$ is not short range in time. Consequently, there is no
Wigner-Weisskopf exponential decay as in (\ref{WW}). While this is true
already for $A=0$, note that the log-periodic modulation of $\Gamma_{e}%
(\omega)$ adds to $\Phi_{e}(t)$ an oscillatory log-periodic term, which
further modifies the behavior of $U_{e}(t)$. To investigate this point in more
detail, we need to study the pole structure in (\ref{persiscorr}). The Laplace
transform of $\Phi_{e}(t)$ is%
\begin{equation}
\tilde{\Phi}_{e}(s-i\omega_{e})={\frac{C}{z^{1-\alpha}}}\left[  \csc{\frac
{\pi\alpha}{2}}+\frac{A}{2i}\left(  \tilde{F}(z)-\tilde{F}^{\ast}(z)\right)
\right]  , \label{phistoy}%
\end{equation}
where $z\equiv s+i\Delta\omega$, $\Delta\omega\equiv\omega_{e}-\omega_{u}$ and
$\tilde{F}(z)\equiv\left(  \frac{z}{\Omega}\right)  ^{2\pi i/\lambda}%
\sinh^{-1}\left(  \frac{\pi^{2}}{\lambda}-\frac{i\pi\alpha}{2}\right)  $. The
poles $s_{n}$ in (\ref{persiscorr}) are solutions of $s+\tilde{\Phi}_{e}%
(z)=0$. For the sake of simplicity, we consider $\Delta\omega=0$, in which
case all the poles come in complex conjugate pairs.
%\begin{figure}[htb]
%\includegraphics[scale=0.25]{pole.pdf}
%\caption{Structure of the poles for the model (\ref{toy}).}
%\label{utoyfig2}
%\end{figure}
For $A$ not too small, namely $A>\csc\left(  \frac{\pi\alpha}{2}\right)
e^{-\pi^{2}/\lambda}$, the poles are infinitely dense near the origin and are
distributed log-periodically with the distance to it.
%(see fig. \ref{poles}).%
In particular, for $\left\vert s_{n}\right\vert ^{2-\alpha}\ll C$, their
expression in the lower half-plane is \cite{ga}%
\begin{equation}
s_{n}\approx-i\Omega\,e^{\lambda\left(  \frac{3-\alpha}{4}+n\right)
+i\theta_{0}}, \label{poles}%
\end{equation}
with $\theta_{0}\equiv\frac{\lambda}{2\pi}\ln\left(  \left\vert A\right\vert
\sin\frac{\pi\alpha}{2}\right)  $ and $n$ is an integer. The corresponding
residues acquire a rather simple form, $\mbox{res}(s_{n})\approx-\sin\left(
\frac{\pi\alpha}{2}\right)  \frac{\lambda}{2\pi iC}s_{n}^{2-\alpha}e^{s_{n}t}%
$. Then, assuming $e^{-2\pi^{2}/\lambda}\ll1$, the branch cut contribution to
(\ref{persiscorr}) can be neglected, and the probability amplitude $\tilde
{U}_{e}(t)\equiv e^{i\omega_{e}t}U_{e}(t)$, in the long time limit $C\left(
\left\vert \sin\theta_{0}\right\vert t\right)  ^{2-\alpha}\,\gg1$, can be
written as \cite{Note-Real-Prob}%
\begin{equation}
\tilde{U}_{e}(t)=-\frac{\lambda\sin\frac{\pi\alpha}{2}}{\pi C}%
\operatorname{Im}\sum_{n=-\infty}^{+\infty}s_{n}^{2-\alpha}e^{s_{n}t}.
\label{utoy}%
\end{equation}
Here, the upper summation limit was extended to $+\infty$ exploiting the
condition on $t$. It is possible to make further approximations and show that
only a few terms in the sum give the main contribution at a given time.
Therefore, locally, the decay envelope appears as a slow exponential, while it
is algebraic at large time scales, $\left\vert U_{e}(t)\right\vert \simeq
\frac{\lambda\sin\left(  \pi\alpha/2\right)  }{C\left(  \left\vert \sin
\theta_{0}\right\vert t\right)  ^{2-\alpha}}$ \cite{ga}. For comparison, the
decay asymptotics for $A=0$ is $\left\vert U_{e}(t)\right\vert \simeq
1/Ct^{2-\alpha}$, as determined by the branch cut in (\ref{persiscorr}). More
interesting is the discrete scaling symmetry displayed in (\ref{utoy}), namely%
\begin{equation}
\tilde{U}_{e}(\beta t)=\beta^{\alpha-2}\,\tilde{U}_{e}%
(t)\,\,\mbox{with}\,\beta=e^{\lambda}, \label{scalingtoy}%
\end{equation}
which is straightforward when using (\ref{poles}) in (\ref{utoy}). More
generally, if for large $t$ one can neglect the free term $s$ in the
denominator in (\ref{persiscorr}), then (\ref{scalingtoy}) immediately follows
from scaling properties of $\tilde{\Phi}_{e}\left(  s-i\omega_{e}\right)  $.
Thus, as a result of (\ref{scaling}), $\tilde{U}_{e}(t)$ takes precisely the
form (\ref{main}) with $\gamma=2-\alpha$. The log-periodic function
$\mathcal{G}$ can be calculated from (\ref{utoy}). The decay amplitude
$U_{e}(t)$ is plotted in Fig. \ref{utoyfig1}. This figure makes explicit that
$U_{e}(t)$ has a much slower decay for $A=1$, than for $A=0$, i.e., without
the log-periodic modulation. As noted above, the decay envelope appears as
exponential since the power law in (\ref{main}) shows up at larger time scales
than displayed in Fig. \ref{utoyfig1}. The oscillating structure of $U_{e}(t)$
shows beats, which result from interferences between the few
%TCIMACRO{\QTR{QQQEvgeni}{dominant} }%
%BeginExpansion
dominant
%EndExpansion
terms in (\ref{utoy}).%
%TCIMACRO{\FRAME{ftbpFU}{2.8908in}{2.1677in}{0pt}{\Qcb{(Color online) Amplitude
%$\tilde{U}_{e}(t)$ as a function of time for the model (\ref{toy}) with
%$\alpha=\frac{1}{2}$, \ $\lambda=1$ and $A=0$ (blue thick line) or $A=1$ (red
%thin line). Inset shows $\left\vert \tilde{U}_{e}(t)\right\vert $ in a
%semi-log scale.}}{\Qlb{utoyfig1}}{utoyinset.eps}%
%width 2.8908in;  height 2.1677in;  depth 0pt;  original-width 5.5625in;
%original-height 4.1702in;  cropleft "0";  croptop "1";  cropright "1";
%cropbottom "0";  filename 'utoyinset.eps';file-properties "XNPEU";}} }%
%BeginExpansion
%EndExpansion

A physical realization of the previous results is provided by an atom, or a
quantum dot, coupled to a vacuum field of a quasi-periodic, \textit{e.g.}
Fibonacci, dielectric cavity \cite{D-Negro-05,Deutch-09}. The latter is made
of a sequence of slabs of two types, $A$ and $B$, of width and refractive
index denoted respectively by $d_{A},d_{B}$ and $n_{A},n_{B}$. Fibonacci
sequences $S_{j}$ of such slabs are constructed by recursion $S_{j\geq
3}=[S_{j-2}S_{j-1}]$, with $S_{1}=B$, $S_{2}=A$. The spectral properties and
the spatial behavior of the corresponding eigenmodes in Fibonacci and some
other quasi-periodic structures have been extensively studied
\cite{QP-Review03}. The spectrum of the Fibonacci system $S_{j}$ is highly
fragmented, with the degree of fragmentation increasing with $j$ and the
contrast $n_{A}/n_{B}$ \cite{Kohmoto-87,Kohmoto-87-Wurtz-88}. More
specifically, the spectral measure $\mu(\omega)$ has a discrete scaling
symmetry (\ref{scaling}) governed by specific $p$-cycles of the associated
renormalization group transformation, whereas $a=\sigma^{p}$ with
$\sigma=\frac{1+\sqrt{5}}{2}$. Near fixed points of a given $p$-cycle map
$T\left(  \omega\right)  $ can be linearized, and $\mu\left(  \omega\right)  $
obeys (\ref{mu}) with $\alpha=\frac{\ln a}{\lambda}=\frac{\ln\sigma^{p}%
}{\lambda}$ and $\lambda=\ln\left\vert T^{\prime}\left(  \omega\right)
\right\vert $. For instance, for $d_{A}n_{A}=d_{B}n_{B}=d$, fixed points of a
$6$-cycle are determined by $\frac{\omega_{n}d}{2\pi c}=\frac{2n+1}{4}$, where
$c$ is the vacuum speed of light and $n=0,1,\ldots$. Linearizing the map at
these points, one obtains $e^{\lambda}=\left\vert T^{\prime}\left(  \omega
_{n}\right)  \right\vert =1+8\eta^{4}+4\eta^{2}\sqrt{1+4\eta^{4}}$ with
$\eta={\frac{1}{2}}\left(  \frac{n_{A}}{n_{B}}+\frac{n_{B}}{n_{A}}\right)  $
\cite{Kohmoto-87-Wurtz-88}. Experimentally, a reasonable contrast can be
achieved with $n_{A}=1.45$ and $n_{B}=2.23$ \cite{D-Negro-05}, so that
$\alpha=0.898$ and $\lambda=3.2$ \cite{Note-Fib-Map}.

In conclusion, we have argued that spontaneous emission of a quantum emitter
coupled to a vacuum characterized by a discrete scaling symmetry exhibits
unusual and measurable features. We have shown that the quantum probability
amplitude never follows the Wigner-Weisskopf decay law (\ref{WW}), but rather
an overall power law whose exponent depends on the fractal properties of the
vacuum. Moreover, we have shown that the decay probability is characterized by
log-periodic fluctuations, both at short and long times. These fluctuations
constitute an unambiguous fingerprint of the underlying fractal structure.
Underneath the slow algebraic decay and log-periodic modulation, the
probability amplitude was shown to exhibit a rich dynamics with nearly
periodic oscillations and beats resulting from the interference between the
main contributions to the pole structure. Finally, we have discussed the
example of a Fibonacci cavity, a non fractal device but whose spectrum is
characterized by a scaling symmetry (\ref{scaling}), and we have argued that
it may be a possible candidate to observe the unusual features of spontaneous
emission described in this letter. Spontaneous emission constitutes a specific
way to probe a fractal nature of spectrum. Thermodynamics \cite{ADT3} and
spatial correlations \cite{ADT2} provide other examples of physical probes.
Let us mention finally, that the approach developed in this letter could be
extended to related problems in quantum mesoscopic physics (Coulomb blockade)
and in the study of quantum vacuum effects such as the dynamical Casimir
effect and the Schwinger and Unruh effects.

This work was supported by the Israel Science Foundation Grant No.924/09.


\begin{thebibliography}{99}
\bibitem {Fract-Coarse-gran}Note that frequencies are dimensionless and
expressed in units of a coarse graining frequency characteristic of the
fractal spectrum.
\bibitem {Note-Ds}In some cases, $\alpha$ is related to the non-local spectral
dimension $d_{s}$, \cite{ADT1}.
\bibitem {Note-IncompDecay}In certain cases, the decaying part (\ref{main})
coexists with a non-decaying term corresponding to the incomplete decay of the
excited state \cite{ga}, similar to that found near the photonic band-edge
\cite{kurizki}.
\bibitem {Note-CouplDOM}We assume that $d\mu(k)|V_{k}|^{2}$\ also obeys the
scaling form (\ref{mu}).
\bibitem {Note-Real-Prob}Note that $\tilde{U}_{e}\left(  t\right)  $ is a real
quantity since the poles $s_{n}$ come in complex conjugate pairs, which is a
consequence of the even parity of the spectral function $\Gamma_{e}\left(
\omega\right)  $ in (\ref{toy}).
\end{thebibliography}
\end{document}